\documentclass[aps, prl, preprint, floatfix]{revtex4-1}
\usepackage{ifthen}
\newboolean{prepr}
\setboolean{prepr}{true}

\usepackage{amsmath,amssymb, amsfonts}
\usepackage{graphicx}

\usepackage[pdftex]{hyperref}

\begin{document}

\title{New relations for scattering amplitudes in Yang-Mills theory at loop level} 
 
\author{Rutger H.  Boels}
\email{Rutger.Boels@desy.de}
\author{Reinke Sven Isermann} 
\email{Reinke.Sven.Isermann@desy.de}
\affiliation{II. Institut f\"ur Theoretische Physik,\\ Universit\"at Hamburg,\\  Luruper Chaussee 149,\\ D-22761 Hamburg, Germany }

\date{\today}



\begin{abstract}
\noindent 
The calculation of scattering amplitudes in Yang-Mills theory at loop level is important for the analysis of background processes at particle colliders as well as our understanding of perturbation theory at the quantum level. We present tools to derive relations for especially one loop amplitudes, as well as several explicit examples for gauge theory coupled to a wide variety of matter. These tools originate in certain scaling behavior of permutation and cyclic sums of Yang-Mills tree amplitudes and loop integrands. In the latter case evidence exists for relations at all loop orders.
\end{abstract}


\maketitle

Most of our understanding of particle physics is based on perturbation theory of the Standard Model in its coupling constants. In terms of textbook Feynman graphs this expansion is correlated with an expansion in terms of loops in the graphs.  Calculating amplitudes using Feynman graphs however becomes rapidly prohibitively complicated even at one loop already. That is not to say that the one loop corrections are not interesting: for various phenomenological processes Next-to-Leading-Order predictions are requested or required \cite{Binoth:2010ra}. The main reason is that the strong nuclear force is strong: its coupling constant is roughly $0.12$ at the mass of the Z boson. Strong quantitative control is needed over this background to gain access to the discovery channels in scattering experiments such at the Large Hadron Collider. 

Motivated by this much effort has been devoted to amplitude calculations in $U(N)$ Yang-Mills theory coupled to matter fields since the Standard Model of particle physics is of this type. Given the complexity of the Feynman graphs surprisingly simple results have been obtained (see e.g. \cite{Dixon:2011xs} and references therein), especially in theories with extended supersymmetry. Some of this simplicity follows from considering symmetry properties properly. For instance, at any loop level one can \cite{Berends:1987cv}  write a scattering amplitude for $n$ gluons as
\ifthenelse{\boolean{prepr}}{
\begin{equation}\label{eq:fullamp}
A^{\textrm{full}}_n = \!\!\! \sum_{\alpha \in P(1,2 \ldots n)\backslash \mathbb{Z}(1,2 \ldots n)} A^{\textrm{co}}(\alpha) \textrm{tr}\left( T^{a_1} T^{a_{2}} \ldots T^{a_{n}} \right) + \textrm{more traces}
\end{equation}}{
\begin{multline}\label{eq:fullamp}
A^{\textrm{full}}_n = \!\!\! \sum_{\alpha \in P(1,2 \ldots n)\backslash \mathbb{Z}(1,2 \ldots n)} A^{\textrm{co}}(\alpha) \textrm{tr}\left( T^{a_1} T^{a_{2}} \ldots T^{a_{n}} \right) \\ + \textrm{more traces}
\end{multline} }
where the sum is over all permutations ($P$) of particles $1$ through $n$ which are not cyclic ($\mathbb{Z}$) and the matrices $T$ are in the fundamental representation of the $U(N)$ group. At $l$ loops there are maximally $l+1$ traces. The so-called color ordered amplitude $A^{\textrm{co}}$ has all particles appearing in a fixed cyclic order.  This decomposition separates the color quantum numbers from kinematics.  The first order in perturbation theory, no loops, is called tree level. All amplitudes in this article will be color-ordered and single trace. 


Color ordering reduces computational complexity as cyclic and inversion symmetry of the trace show there are maximally $(n-1)!/2$ independent color ordered amplitudes, versus $n!$ different full ones. Further relations are known \cite{Kleiss:1988ne} which reduce this number to $(n-2)!$. Recently it was shown by Bern, Carrasco and Johansson (BCJ)   \cite{Bern:2008qj} there are additional relations which yield  $(n-3)!$ independent amplitudes. These show tree level amplitudes are much simpler than previously thought.  

At the one loop level less is known. There are relations \cite{Bern:1994zx}  which relate the double trace terms to the single trace in equation \eqref{eq:fullamp}. Further relations known or conjectured are restricted to special choices of helicities \cite{Bern:1993qk} \cite{BjerrumBohr:2011xe}: either all equal or one unequal for all gluons in the amplitude. With all helicities equal for instance a so-called ``three photon decoupling identity'',
\begin{equation}\label{eq:vanamp1}
\sum_{\sigma \in POP(\alpha_3 \cup \beta) }A^{1-\textrm{loop}}(\sigma) = 0\quad \textrm{     (helicity equal)}
\end{equation}
holds  \cite{Bern:1993qk} with the sum over all partially ordered products (POP): all unions of the sets $\alpha$ and $\beta$ keeping the order of the set $\beta$ fixed. Numerical evidence was found in \cite{BjerrumBohr:2011xe} that these relations also apply to the one-helicity unequal case. The sub-index of the set $\alpha$ indicates the number of particles in the set. This sum arises by setting the color matrices corresponding to the three photons in the set $\alpha$ to the identity in \eqref{eq:fullamp} and collecting terms. It is the purpose of this article to introduce methods to obtain helicity blind relations at one loop and possibly beyond. 

For general helicities the analytic structure of one-loop color-ordered Yang-Mills amplitudes can be captured in a standard basis of scalar integrals, (see e.g. \cite{Dixon:1996wi})
\ifthenelse{\boolean{prepr}}{
\begin{equation}\label{eq:massiveloopampexp}
A_n^{\textrm{co}, 1-\textrm{loop}} = \sum a_b \left(\textrm{Boxes} \right) +  a_t  \left(\textrm{Triangles} \right)  + a_{bb}  \left(\textrm{Bubbles} \right) +  \textrm{Rational}
\end{equation} }{
\begin{multline}\label{eq:massiveloopampexp}
A_n^{\textrm{co}, 1-\textrm{loop}} = \sum a_b \left(\textrm{Boxes} \right) +  a_t  \left(\textrm{Triangles} \right) \\ + a_{bb}  \left(\textrm{Bubbles} \right) +  \textrm{Rational}
\end{multline}}where the last terms are simply rational functions of polarizations and momenta. The sum ranges over all ``channels'': all ways to distribute the external particles over the corners of the integrals, leaving no corner empty. As the integrals can be integrated once and for all so the problem of calculating one loop amplitudes reduces to obtaining the integral coefficients $a_i$ which are rational functions of the external momenta and polarizations. These coefficients can be expressed as functions of tree level amplitudes \cite{Britto:2004nc, Forde:2007mi, Badger:2008cm} with the box coefficients the simplest and the rational terms the most complicated. 


Special  theories are known to have vanishing coefficients in the expansion of \eqref{eq:massiveloopampexp}. Supersymmetric gauge theories with massless particles generically do not have rational terms  \cite{Bern:1994cg}. Maximally supersymmetric gauge theories in four dimensions have vanishing bubble and triangle coefficients  \cite{Bern:1994zx}, which makes the loop amplitudes here simple to calculate. Similar results may be obtained in the theory of Quantum Electrodynamics (QED) \cite{Badger:2008rn}. One method to generate relations for coefficients in the expansion \eqref{eq:massiveloopampexp} below follows from this QED example. A related method suggests relations for the integrand of one loop amplitudes which reduce the number of independent one loop integrands  from $(n-1)!/2$ to $(n-2)!$. All of these show Yang-Mills theories at loop level are simpler than previously thought. 

Both methods are rooted in the analysis of scaling behavior of certain permutation and cyclic sums over Yang-Mills tree amplitudes and integrands at loop level. These translates through the formulae of \cite{Britto:2004nc, Forde:2007mi, Badger:2008cm} to statements about one loop amplitude coefficients and through on-shell recursion relations  \cite{Britto:2004ap, Britto:2005fq, ArkaniHamed:2010kv, Boels:2010nw} to the integrand. The results of this analysis will be presented first.

\section{Scaling of permutation and cyclic sums of tree amplitudes}\label{sec:trees}
The blue-print for the scaling behavior is the analysis of large so-called Britto-Cachazo-Feng-Witten (BCFW) \cite{Britto:2004nc, Britto:2004ap} shifts for Yang-Mills theory tree amplitudes in four and higher dimensions. The BCFW shift changes the momenta of two gluons as
\begin{equation}\label{eq:bcfwshift}
k^{\mu}_i \rightarrow k^{\mu}_i + q^{\mu} z \quad k^{\mu}_j \rightarrow k^{\mu}_j - q^{\mu} z
\end{equation}
with $q$ constrained to obey $q \cdot k_i = q \cdot k_j = q \cdot q = 0$. It can be shown (see e.g. \cite{ArkaniHamed:2008yf}) that for large $z$ a tree amplitude scales under the shift of color-adjacent particles, say $i$ and $i+1$, as  
\begin{equation}
A_n(1,\ldots, i, i+1, \ldots n) \rightarrow  \xi^i_{\mu} \xi^{i+1}_{\nu} G^{\mu\nu}(z) 
\end{equation}
where the functional form $G$ is given by 
\begin{equation}
G^{\mu\nu}(z) \equiv \left(z \, \eta^{\mu\nu} f(1/z) + B^{\mu\nu}(1/z) + \mathcal{O} \left( \frac{1}{z}\right) \right)
\end{equation}
The $\xi$'s are the $z$-dependent polarization vectors of the shifted legs, $f$ and $B^{\mu\nu}$ are polynomial functions of $1/z$ whose coefficients are unimportant here and $B^{\mu\nu}$ is anti-symmetric. This result is crucial for proving on-shell recursion relations \cite{Britto:2004ap}. Generic BCFW shifts of particles which are not color-adjacent are known to display better, $1/z$-suppressed scaling behavior for tree amplitudes. In QED even more improved scaling is related to the vanishing of rational, bubble and triangle terms \cite{Badger:2008rn}. 

We conjecture two closely related ways to improve BCFW scaling of tree amplitudes in Yang-Mills theory. The first is that the shift of particles $i$ and $j$ on either side of a permutation sum scales as 
\begin{equation}\label{eq:permconj}
\sum_{\alpha \in P(\{i+1 \ldots j-1\})}  A_n(\ldots i, \alpha, j, \ldots)  \rightarrow  \xi^i_{\mu} \xi^{j}_{\nu} \frac{G^{\mu\nu}(z) }{z^{j-i-1}}
\end{equation}
as long as $i$ and $j$ are not color adjacent (i.e. the dots contain at least one particle leg). If they are, $1/z^{j-i-1} \rightarrow 1/z^{j-i-2}$. The second is that the shift of particles $i$ and $j$ on either side of a cyclic sum scales as 
\begin{equation}\label{eq:cyclconj}
\sum_{\alpha \in \mathbb{Z}(\{i+1 \ldots j-1\})}  A_n(\ldots i, \alpha, j, \ldots)  \rightarrow  \xi^i_{\mu} \xi^{j}_{\nu}  \frac{G^{\mu\nu}(z) }{z^2}
\end{equation}
as long as $i$ and $j$ are not color adjacent. If they are, $1/z^2 \rightarrow 1/z$. These conjectures can be combined to give
\begin{equation}
\sum_{\substack{ \alpha \in   \mathbb{Z}(\{i+1 \ldots j-1\}) \\ \beta \in P(\{j+1 \ldots i-1\}) }}  A_n(\alpha, i, \beta, j) \rightarrow  \xi^i_{\mu} \xi^{j}_{\nu} \frac{G^{\mu\nu}(z) }{z^{j-i}}
\end{equation}
for non-empty sums.  Proof of these conjectures at tree level up to and including $1/z^2$ terms is provided in \cite{companion} by adapting the techniques of \cite{ArkaniHamed:2008yf}, as well as a proof of the generic non-adjacent ($j=i+2$ in equation \eqref{eq:permconj}) shift for the integrand of Yang-Mills amplitudes at any loop order by extending \cite{Boels:2010nw}. There is a natural extension to shifts in theories with matter coupled to glue. The case of shifts of massive scalar legs is treated in  \cite{companion} as well: the same improvement in shift behavior as displayed here for gluons is found. 

These conjectures can at tree level be extended to other amplitudes by use of the Kleiss-Kuijf  \cite{Kleiss:1988ne} relations, 
\begin{equation}\label{eq:KKrel}
A_n (1,\alpha,2,\beta) =  (-1)^{\# \alpha} \sum_{\omega \in OP(\alpha^T \cup \beta)} A_n(1, 2, \omega)
\end{equation}
where the ordered product (OP) is the set of all unions of the sets $\alpha^T$ and $\beta$ keeping the order of both the subsets fixed. The transpose on $\alpha^T$ inverts the order of the set. 
These scaling results will be employed below to derive identities for scattering amplitudes at one loop. 

\section{Three photon decoupling relation for rational terms}\label{sec:loopsI}
It will be shown here that  equation \eqref{eq:vanamp1} generalizes to the rational part of any 1-loop amplitude for more than four points, i.e.
\begin{equation}\label{eq:vanratis}
\textrm{Rational} \left[ \sum_{\sigma \in POP(\alpha_3 \cup \beta) }A_n^{1-\textrm{loop}}(\sigma) \right]= 0
\end{equation}
The first non-trivial example of these relations is at six points. They have been numerically cross-checked \cite{badgerprivate} up to seven points using NGluon \cite{Badger:2010nx}.

\begin{figure}[h!]
\centering
\includegraphics[scale=0.6]{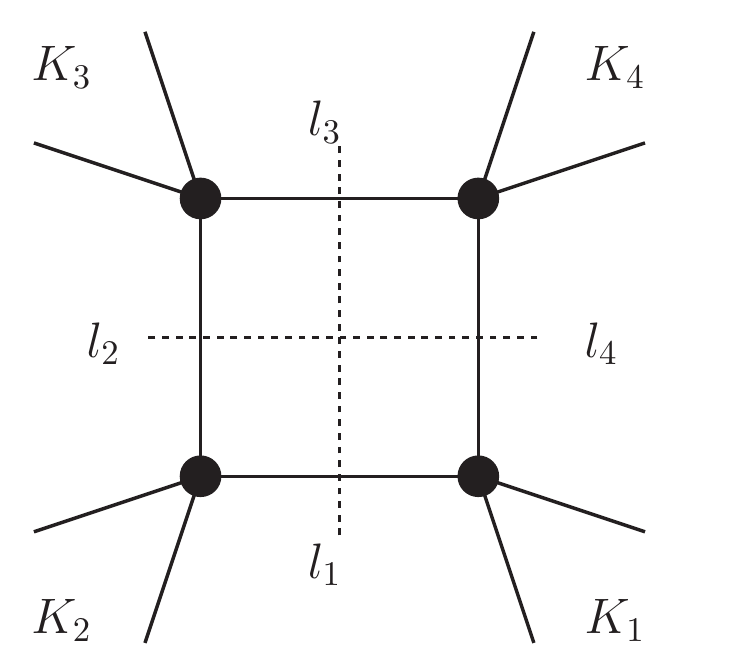}
\caption{\label{fig:quadruplecut} Momentum assignment of a massive box coefficient. The $i$-th corner contains gluons with total momentum $K_i$.  }
\end{figure}

The rational terms can  \cite{Badger:2008cm} be written similar to equation \eqref{eq:massiveloopampexp} as
\ifthenelse{\boolean{prepr}}{
\begin{equation}
\textrm{Rational} = \sum \left(\textrm{massive Box} \right) +  \left(\textrm{massive Triangle} \right ) + \left(\textrm{massive Bubbles} \right) 
\end{equation} }{
\begin{multline}
\textrm{Rational} = \sum \left(\textrm{massive Box} \right) +  \left(\textrm{massive Triangle} \right )\\ + \left(\textrm{massive Bubbles} \right) 
\end{multline}}
The massive box term is illustrated in figure \ref{fig:quadruplecut}. The particle in the loop is a massive scalar with mass $\mu^2$. The corners are formed by amplitudes of gluons coupling to a massive scalar pair and the sum ranges over all channels. In a given channel the massive box coefficient reads \cite{Badger:2008cm}
\ifthenelse{\boolean{prepr}}{\begin{multline}\label{eq:cutmassivebox}
\left(\textrm{massive Box} \right) =  \\ \sum_{\sigma} \textrm{Inf}_{\mu}  \left[A_{\phi \bar{\phi}}(-l_1, K_1, l_2) A_{\phi \bar{\phi}}(-l_2, K_2, l_3) A_{\phi \bar{\phi}}(-l_3, K_3, l_4)A_{\phi \bar{\phi}}(-l_4, K_4, l_1) \right]_{\mu^4} 
\end{multline}}{
\begin{multline}\label{eq:cutmassivebox}
\left(\textrm{massive Box} \right) =\sum_{\sigma} \textrm{Inf}_{\mu}  \left[A(-l_1, K_1, l_2)  \right. \\ \left. A(-l_2, K_2, l_3) A(-l_3, K_3, l_4)  A(-l_4, K_4, l_1) \right]_{\mu^4} 
\end{multline}
}
The first and last entry on the amplitude correspond to the scalar momenta, while capital $K_i$ are the sum of all incoming gluon momenta. The symbol $\textrm{Inf}_x$ stands for taking the polynomial part of the Laurent series of a function $f(x)$ at $x \rightarrow \infty$ and the final instruction is to isolate the $\mu^4$ term in this polynomial. All loop momenta are on-shell, $l_i^2 = \mu^2$. These constraints can be solved as
\begin{equation}\label{eq:solsforthreeconstr}
l_{1,\nu} = a K_{1,\nu}^{\flat} + b K_{2,\nu}^{\flat} + t E_{\nu} + \frac{\gamma_{14} ab - \mu^2}{t \gamma_{14}} \bar{E}_{\nu} 
\end{equation}
where $K_1^{\flat}$ and $K_2^{\flat}$ are light-like and defined by
\begin{align}
K_{1,\nu} & = K_{1,\nu}^{\flat} + \frac{K_1^2}{2 K^{\flat}_1 \cdot K^{\flat}_2} K^{\flat}_{2,\nu} \qquad (1 \leftrightarrow 2) 
\end{align}
Vectors $E$ and $\bar{E}$ are light-like, orthogonal to $K_1$ and $K_2$ and normalized to $ K_1^{\flat} \cdot K_2^{\flat} = - E \cdot \bar{E} \equiv \gamma_{14}$. Hence $l_1^2 = \mu^2$. The constants $a$ and $b$ follow from $l_2^2 = \mu^2$ and $l_4^2 = \mu^2$. Finally, $t$ are the two solutions to $(l_3)^2 = \mu^2$. These two are summed over in equation \eqref{eq:cutmassivebox}. In the limit $\mu \rightarrow \infty$ all loop momenta scale as
\begin{equation}
l^{\nu}_i \rightarrow \mu V^{\nu} + \mathcal{O}\left( \mu^0\right)
\end{equation}
for the space-like vector $V^{\nu}$ which is orthogonal to all external momenta $K^{\nu}_i$. There is a close relation between this and the BCFW shift \eqref{eq:bcfwshift} and a very similar analysis applies. Generically all amplitudes will scale as $\mu$ for instance, while permutation or cyclic sums improve the scaling behavior completely analogous to the BCFW shift cases in equations \eqref{eq:permconj} and \eqref{eq:cyclconj}. 

\begin{figure}[h]
\centering
\includegraphics[scale=0.5]{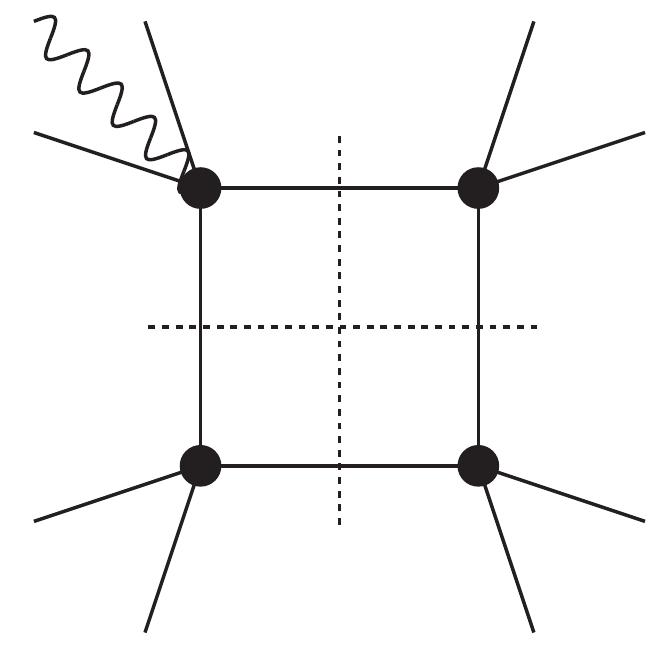}
\caption{\label{fig:onephotonmassivebox} Massive box coefficient for an amplitude with one photon}
\end{figure}
\begin{figure}[h!]
\centering
\includegraphics[scale=0.4]{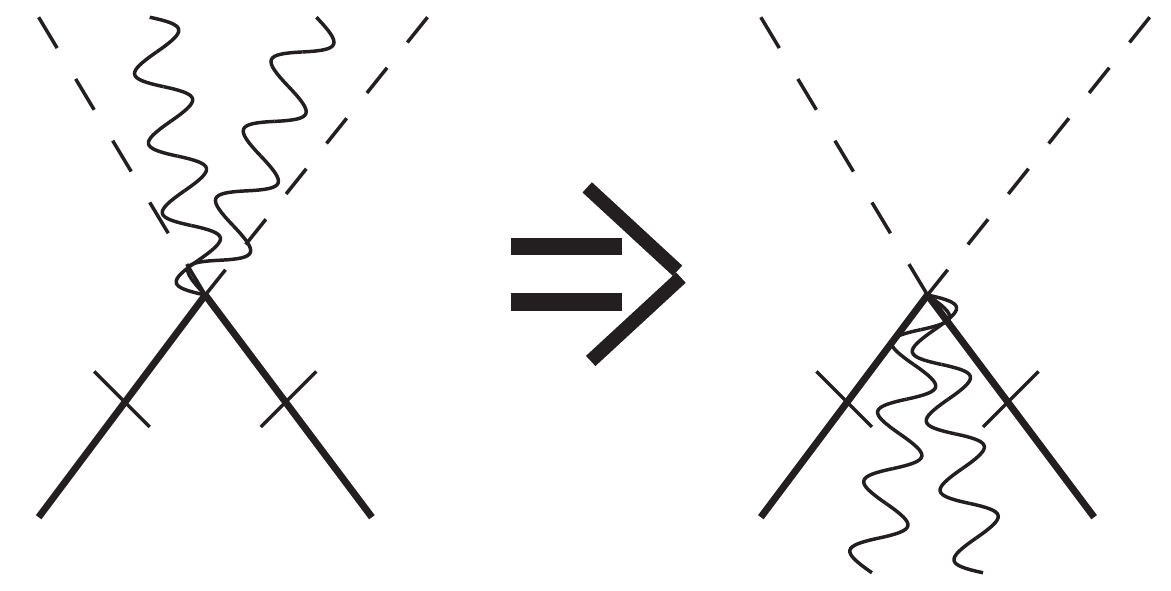}
\caption{\label{fig:useofKK} The Kleiss-Kuijf relation (\eqref{eq:KKrel}) relates adjacent shifts of massive scalars coupled to photons and gluons to non-adjacent shifts}
\end{figure}

Now consider turning one gluon into a photon. The massive box coefficients in a generic channel will always involve an amplitude where one photon is present in addition to gluons, see figure \ref{fig:onephotonmassivebox}. This photon appears in all possible positions w.r.t. the gluons. Hence using the Kleiss-Kuijf relation the photon can be moved to the other side of the shifted scalar legs, see figure \ref{fig:useofKK}. Since this will improve the scaling of the amplitude \cite{companion}, the resulting massive box coefficient in equation \eqref{eq:cutmassivebox} will vanish in this generic channel. An exception to this occurs when the photon couples to a corner of the box without gluons since in this case no improvement in scaling results.

The previous argument can be repeated for multiple photons. In a generic channel the permutation sum of \eqref{eq:vanratis} will make the corresponding massive box coefficient vanish, while through the massive scalar equivalent of \eqref{eq:permconj} exceptions occur only when single photons couple to a corner. Hence for four photons, as long as there is one additional gluon, the massive box coefficients always vanish. In the boundary case of three photons the gluons have to couple to a single corner. However, in this case a cyclic sum over the gluons arises which by equation \eqref{eq:cyclconj} makes this contribution also vanish. Note the latter argument fails in the four point case ($n=4$ in equation \eqref{eq:vanratis}).

A similar analysis also applies to the massive triangle and massive bubble coefficients \cite{companion} which concludes the proof of \eqref{eq:vanratis}.

\section{Five photon decoupling relation for triangle and bubble terms}\label{sec:loopsII}
The bubble and triangle terms in equation \eqref{eq:massiveloopampexp} can be shown to cancel in the following sum for more than six particles,
\begin{equation}\label{eq:vanbubtri}
\begin{array}{c}
\textrm{Bubble} \\ \textrm{or} \\
\textrm{Triangle} \end{array} \left[ \sum_{\sigma \in POP(\alpha_5 \cup \beta) }A_n^{1-\textrm{loop}}(\sigma) \right]  = 0
\end{equation}
Combined with equation \eqref{eq:vanratis} this shows that for five ``photons'' only the simple box terms remain in equation \eqref{eq:massiveloopampexp}. The first non-trivial example of these relations is at eight points.  They have been numerically cross-checked \cite{badgerprivate} at eight points using NGluon \cite{Badger:2010nx}.

The proof of \eqref{eq:vanbubtri} is based on the formulae of  \cite{Forde:2007mi} which express the coefficients in terms of tree level four dimensional Yang-Mills amplitudes. For triangle coefficients for instance
\ifthenelse{\boolean{prepr}}{ \begin{equation}\label{eq:trianglecontr}
 \left(\textrm{Triangle} \right) = \sum_{\omega} \textrm{Inf}_{t} \sum_{\textrm{helicities}}  \left(A(-l_1, K_1, l_2) A(-l_2, K_2, l_3) A(-l_3, K_3, l_1) \right)_{t=0}
\end{equation}}{
\begin{multline}\label{eq:trianglecontr}
 \left(\textrm{Triangle} \right) = \sum_{\omega} \textrm{Inf}_{t} \sum_{\textrm{helicities}} \\ \left(A(-l_1, K_1, l_2) A(-l_2, K_2, l_3) A(-l_3, K_3, l_1) \right)_{t=0}
\end{multline}}
with the loop momentum parametrized as in \eqref{eq:solsforthreeconstr} with $\mu^2= 0$ and $t$ unconstrained. New here is the sum over the helicities of the on-shell four dimensional gluon in the loop. This sum leads to the completeness relation
\begin{equation}\label{eq:complrel}
 \sum_{\textrm{helicities}} \xi_{\mu}(l) \xi_{\nu}(l) = \eta_{\mu\nu} - \frac{n_{\mu} l_{\nu} + n_{\nu} l_{\mu}}{l \cdot n}
\end{equation}
for polarization vectors with $n_{\mu}$ a gauge choice. In the limit $t \rightarrow \infty $ the right hand side of equation \eqref{eq:complrel} scales as $t^0 + \mathcal{O}(\frac{1}{t})$ and has a universal limit for all three cut legs. This leaves the scaling of the tree amplitudes in \eqref{eq:trianglecontr}.  For five photons the only remaining terms after applying \eqref{eq:permconj} vanish by  \eqref{eq:cyclconj}, unless there is only one gluon (which corresponds to the six particle case). Note these vanishing results are based on permutation and cyclic sums only so apply to the coefficient of the triangle integral in every channel seperately. The analysis of the bubble coefficients is similar which completes the proof of \eqref{eq:vanbubtri}. Note that for less ``photons'' in equations \eqref{eq:vanratis} or  \eqref{eq:vanbubtri}  simplifications occur, but terms remain. 

\section{Relations for massive box contributions to rational terms}\label{sec:loopsIII}
Recently novel relations for helicity equal amplitudes at one loop have been conjectured in \cite{BjerrumBohr:2011xe} and verified for low-multiplicity cases. These have a structural similarity to equation  \eqref{eq:vanamp1} and can be written as 
\ifthenelse{\boolean{prepr}}{\begin{equation}\label{eq:312bjbspec}
\sum_{P(4,\ldots n)} \left[ A^{1-\textrm{loop}}_n(1 4 2 3 5 \ldots n) +  A^{1-\textrm{loop}}_n(1 2 4 3 5 \ldots n) + (n-6) A^{1-\textrm{loop}}_n(1 2 3 4 5 \ldots n) \right] = 0 \nonumber
\end{equation}\vspace{-0.6cm}}{
\begin{multline}\label{eq:312bjbspec}
\sum_{P(4,\ldots n)} \!\!\! \left[ A^{1-\textrm{loop}}_n(1 4 2 3 5 \ldots n) +  A^{1-\textrm{loop}}_n(1 2 4 3 5 \ldots n) + \right. \\  (n-6) A^{1-\textrm{loop}}_n(1 2 3 4 5 \ldots n) \big] = 0 \nonumber
\end{multline}\vspace{-0.6cm}
}
\begin{equation}\label{eq:316bjb}
6 \, A^{1-\textrm{loop}}_n(1,2,\ldots, n) - \sum_{k=2}^{n-1} \sum_{\sigma \in OP(\alpha_k \cup \beta_k)} \!\!\!\!\! \left[ A^{1-\textrm{loop}}_n(1, \sigma) \right] =0 \nonumber
\end{equation}
with the sets $\alpha_k  = \{2,\ldots, k\} $ and $\beta_k   = \{k+1,\ldots, n\}$ ordered sequences of gluons. Helicity equal amplitudes only have massive box contributions in equation \eqref{eq:cutmassivebox}. Using equation \eqref{eq:permconj} and \eqref{eq:cyclconj}  it can be shown \cite{companion} that the written relations hold for the massive box coefficients of the written amplitudes independent of helicity. The minimal basis of massive box coefficients modulo the relations generated by  equations \eqref{eq:permconj} and \eqref{eq:cyclconj} can be shown \cite{companion} to contain $|S_1|^{n-1}_3$ elements, where $S_1$ is the Stirling number of the first kind.

\section{Relations for the Yang-Mills integrand at one loop}
At tree level the scaling of equation \eqref{eq:permconj} for one particle in the permutation sum and the on-shell recursion relations of  \cite{Britto:2004ap, Britto:2005fq} have been used to prove \cite{Feng:2010my} the BCJ relations between tree amplitudes in Yang-Mills theory conjectured in \cite{Bern:2008qj}. Similar on-shell recursion relations for the Yang-Mills integrand have been investigated in \cite{ArkaniHamed:2010kv} and \cite{Boels:2010nw}. Hence the question is if the derivation of BCJ type relations of \cite{Feng:2010my} generalizes to the integrand case as well, since it can be shown \cite{companion} the scaling property of a generic non-adjacent shift of the integrand obeys equation \eqref{eq:permconj} (for $j-i=2$) to all loop orders.

\begin{figure}[h!]
\centering
\includegraphics[scale=0.4]{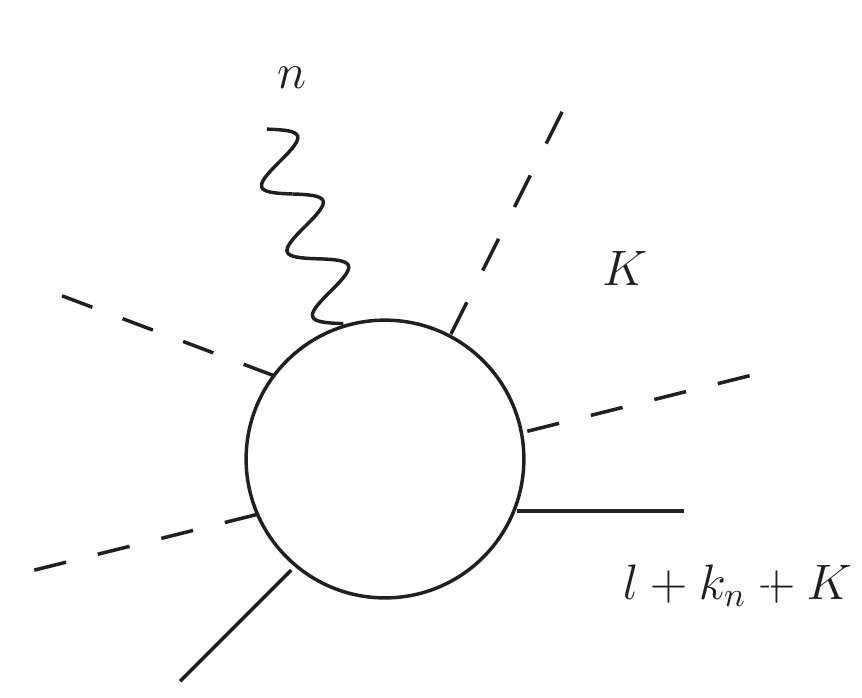}
\caption{\label{fig:loopmomconv} Convention for choosing loop momenta. The bubble represents any color ordered Feynman graph for the planar amplitude. The solid lines indicate the loop legs.}
\end{figure}

At one loop we have experimented with the on-shell recursion relations to find a generalization of the BCJ relations to the color ordered one loop integrand. The recursion relations suggest a choice of convention for the loop momentum: the loop propagator after the point where particle $n$ attaches to the loop is of the form $1/(l+k_{n-1} + K)^2$, where
$K$ is the sum of momenta of all particles between $n$ and the loop propagator in the color ordered graph. This choice which is  illustrated in figure \eqref{fig:loopmomconv} fixes the loop momentum uniquely. With this choice one obtains
\begin{equation}\label{eq:BCJatoneloop}
\sum_{i=2}^{n} k_1 \cdot (l+k_{n} + \sum_{j=2}^{i-1} k_j) I_{n}(2\ldots i-1,1,i,\ldots n) = 0
\end{equation}
as a relation for the integrand in any renormalizable gauge theory with adjoint matter on the external legs, including pure Yang-Mills. The zero here is up to terms which vanish after integration.

If the so-called single-cut contribution is given by a naive extension of the suggestion in \cite{CaronHuot:2010zt} the above relation can be shown to hold in supersymmetric gauge theories using on-shell recursion. This argument leads however quite far from the scope of the present article. Independent of supersymmetry it is straightforward to show \cite{companion} this combination of integrands does not have any two particle unitarity cuts in $D$ dimensions. Since all singularities of \eqref{eq:BCJatoneloop} follow by Cutkosky's rules \cite{Cutkosky:1960sp} this proves the relation: after integration this function vanishes. This proves the relation. Relations of the type \eqref{eq:BCJatoneloop} can be iterated to reduce the number of independent integrands from $(n-1)!/2$ to $(n-2)!$, assuming no accidental degeneracies of the coefficients.

\section{Conclusion and discussion}
This article initiates the systematic study of relations for loop amplitudes and integrands in Yang-Mills theories, including a variety of new examples. Generating an exhaustive list of relations by the methods of this article as well as minimal solution sets would be very interesting. Similar relations for amplitudes and integrands in QCD should exist in which contain external quark lines. The case of amplitudes with quarks in the loop can be related by supersymmetric decomposition of the particles in the loop (see e.g. \cite{Dixon:1996wi}) to the cases of either a gluon or a scalar in the loop contained in the discussion above. 

Further interesting questions surround the possible applications of coefficient relations to streamline numerical calculations. The calculation of rational terms as well as the calculation of the double trace terms of equation \eqref{eq:fullamp} are bottlenecks in the numerical evaluation of one loop amplitudes. Utilizing and extending the relations found for massive boxes is therefore a priority. Also, by the formulae in \cite{Bern:1994zx}  the sums in equations  \eqref{eq:vanratis} and \eqref{eq:vanbubtri} can be related to certain sums over color ordered double trace amplitudes. A better understanding of both these points is desirable. 

Further investigation of equation \eqref{eq:BCJatoneloop} should be very interesting.  How equation \eqref{eq:BCJatoneloop} interacts with the standard expansion of equation \eqref{eq:massiveloopampexp} for instance is a first goal. From the proof \cite{companion} of improved generic non-adjacent BCFW shifts to all loop order for the integrand it is expected generalizations of relations of this type to higher loop integrands exist in principle. This is also suggested by the extension of the conjectured organization of amplitudes in \cite{Bern:2008qj} which lead to the original tree level BCJ relations to loop integrands in \cite{Bern:2010ue}. The higher loop relations would show gauge theories are also at loop level much simpler than previously thought. 

\section*{Acknowledgements}
We would like to thank Simon Badger, Zvi Bern, Emil Bjerrum-Bohr and Michael Kiermaier for discussions. This work has been supported by the German Science Foundation (DFG) within the Collaborative Research Center 676 ``Particles, Strings and the Early Universe". 

\appendix

\bibliographystyle{apsrev4-1}

\bibliography{../NonAdjacentShifts/nonadjbib}

\end{document}